\def\cm2{cm$^{-2}$}

\def\c2{C~{\sc ii}}
\def\c4{C~{\sc iv}}
\def\fe2{Fe~{\sc ii}}
\def\fe3{Fe~{\sc iii}}
\def\mg1{Mg~{\sc i}}
\def\mg2{Mg~{\sc ii}}
\def\si2{Si~{\sc ii}}
\def\si4{Si~{\sc iv}}
\def\al2{Al~{\sc ii}}
\def\al3{Al~{\sc iii}}
\def\o1{O~{\sc i}}
\def\n1{N~{\sc i}}
\def\h1{H~{\sc i}}

\def\approxlt{\mathrel{\spose{\lower 3pt\hbox{$\sim$}}
        \raise 2.0pt\hbox{$<$}}}
\def\approxgt{\mathrel{\spose{\lower 3pt\hbox{$\sim$}}
        \raise 2.0pt\hbox{$>$}}}

\documentclass[article]{has40}
\usepackage{graphicx}
\usepackage{amssymb}
\tabletypesize{\normalsize}  % AMcW

\def\plotone#1{\centering \leavevmode
\includegraphics[width=.95\columnwidth]{#1}}

\def\plotone#1{\centering \leavevmode
\includegraphics[width=.95\columnwidth]{#1}}

\shortauthors{Steenwyk et al.}
\shorttitle{Atypical Light Curves}

\begin{document}
\large    %AMcW  The conference proceedings will employ large size print
\pagenumbering{arabic}
\setcounter{page}{42}

\title{Atypical Light Curves in Close Binaries}

\author{{\noindent Steven D. Steenwyk{$^{\rm 1}$}, Daniel M. Van Noord{$^{\rm 1}$}  and Lawrence A. Molnar{$^{\rm 1}$} \\
\\
{\it (1) Department of Physics and Astronomy, Calvin College, Grand Rapids, MI, USA\\
} 
}
}

\email{(1) ssteen@calvin.edu}

\begin{abstract}
We have identified some two-hundred new variable stars in a systematic study of a data archive obtained with the Calvin-Rehoboth observatory. Of these, we present five close binaries showing behaviors presumably due to star spots or other magnetic activity. For context, we first present two new RS CVn systems whose behavior can be readily attribute to star spots. Then we present three new close binary systems that are rather atypical, with light curves that are changing over time in ways not easily understood in terms of star spot activity generally associated with magnetically active binary systems called RS CVn systems. Two of these three are contact binaries that exhibit gradual changes in  average brightness without noticeable changes in light curve shape.  A third system has shown such large changes in light curve morphology that we speculate this may be a rare instance of a system that transitions back and forth between contact and noncontact configurations, perhaps driven by magnetic cycles in at least one member of the binary. 
\end{abstract}

\section{INTRODUCTION}
Discovery and characterization of close binary systems, especially contact binaries of W Ursae Majoris type, has been the primary focus of variable star observations by our group using the 0.4 m telescopes of the Calvin College Observatory situated in Rehoboth, New Mexico and on the College's main campus in Grand Rapids, Michigan. Searching our own archival data, we discovered five close binary systems whose light curves show dynamic changes that deviate to varying degrees from the simple repetitive forms that characterize typical members of their binary class.  Though we lack spectroscopic evidence of chromospheric activity, we are confident that magnetic activity is the root cause of the dynamic light curve changes for at least three of the five cases.  
  
The connection between magnetic activity on the Sun and sunspots was published by George Ellery Hale in 1908.  Since then, the existence of stellar spots and associated magnetic activity has become an essential element in astronomers' understanding of stars and their observed variability.   Magnetic fields in rotating stars with convective envelopes are generally well understood by so-called stellar dynamo models.  Broadly, the electromagnetic  coupling of rotational and convective motions in a star's electrically conductive envelope generates localized regions of magnetic field lines that emerge from the photosphere and extend into the chromosphere and corona before reentering the photosphere and completing closed magnetic loops in the interior.  The fields suppress upward convective motion of hot plasma, thus locally lowering the surface temperature to create dark spots in the photosphere.  This magnetic activity can also generate bright flares.  RS CVn systems display chromospheric  activity, meaning they emit spectral lines from their chromospheres that are diagnostic of relatively strong magnetic fields.  If simply carried along with the star's rotation, star spots will create repetitive variations in brightness in synch with its rotational period.  In close binaries these variations will be superposed on the periodic brightness variations due to eclipses and tidal shape distortions.  However, spots are generally dynamic features showing changes in their number, shape and location on a star that will lead to nonrepetitive changes in light curve shape.  In close binaries, the rapid rotation, with periods of hours to a few days, is believed to be responsible for especially strong and extensive magnetic fields generating correspondingly large star spots that may cover significant portions of the stellar disk, producing very noticeable distortions in light curve shape that may change on time scales ranging from days to months.  

\section{TWO TYPICAL RS CVn BINARIES}
RS Canum Venaticorum stars are close binaries, normally detached, but only a fraction of close binaries are RS CVn types.  In that sense they are atypical types of close binaries, whether detached or in contact.  Figure 1 shows a light curve of a magnetically active binary classified as EA/RS when submitted to the Variable Star Index (VSX J205514.1+083328) of the American Association of Variable Star Observers (AAVSO).  The EA code is given because the two distinct eclipses show it is a detached binary.  The RS designation arises from apparent star spot activity. It is clear from the June 2010 data (red dots) that, the light curve is significantly lower coming out of the secondary eclipse as well as in the center near phase 0.25, indicating the presence of large star spots suppressing the light output.  However, by August (dark yellow points) the spot that was evident between phase 0.5 and 0.6 has largely disappeared though there remains a generally lower brightness over that region compared to the data from the summer of 2011.  Unfortunately, the August 2010 data do not cover the flat region around phase 0.25 to follow the evolution of that spot.  The region around phase 0.55 shows a total change of around 0.06 magnitudes that is fairly typical for such RS CVn systems, though variation in excess of 0.2 magnitude has been reported in other systems.  

\begin{figure*}
\centering
\includegraphics[width=12cm]{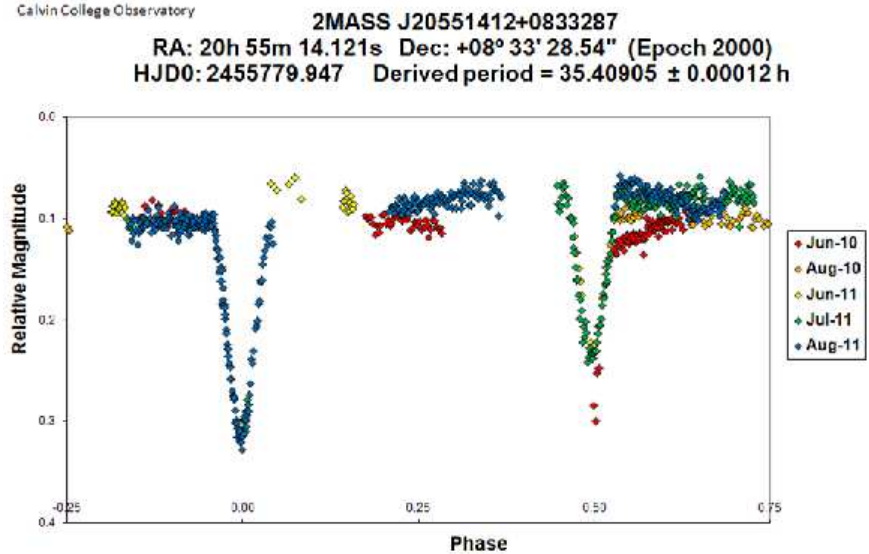}
%\plotone{Steenwyk_Fig1a.eps}
\vskip0pt
\caption{Newly discovered eclipsing RS-CVn, a classic example }
\label{o1}
\end{figure*}

Figure 2 shows the second RS CVn system (VSX J154427.0$-$184425). With a short period of just over three days, it is unlikely to be a single star. However, it shows no evidence of eclipse, so a binary type could not be specified. The 2011 data show a brightness drop of about 0.03 magnitude around phase 0.6 suggesting the occurrence of at least one large stellar spot expressed at that phase that was not present in the 2010 data.  Differential rotation on the chromospherically active star causes spots at different latitudes to move at different rates across the stellar disk. That produces a modulation in the light curve at a slightly different period than the orbital period.  This behavior has been characteristic of many eclipsing RS CVn systems and may well be present here.

\begin{figure*}
\centering
\includegraphics[width=12cm]{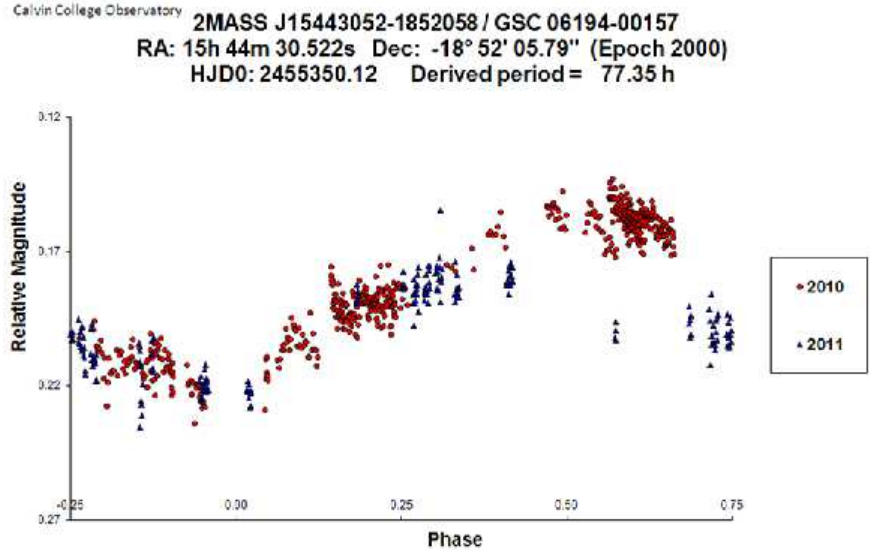}
%\plotone{Steenwyk_Fig2.eps}
\vskip0pt
\caption{Newly discovered non-eclipsing RS-CVn}
\label{o2}
\end{figure*}

\section{TWO STARS WITH SLOW LUMINOSITY CHANGES}
We have found two contact binaries showing a slow shift in average brightness with no evidence of changing light curve shape. In three successive nights in 2012 we obtained the near textbook light curve shown by green and yellow diamonds in Figure 3 for a W UMa type contact binary (VSX J133737.8+084245).  However, it is clear the light curve has shifted downward in brightness compared to the red triangles showing  discovery data obtained from our 2007 archives.  Upon seeing the shift, we sought additional measurements of this star in the data released by the Catalina Real-time Transient Survey, or CRTS (Drake 2009). These data provide a long term record of its brightness over the years 2005-2012, albeit more sparsely and with lower signal-to-noise. We are confident the observed shift is not an artifact of differences in calibration since we recalibrated the CRTS data using the same five reference stars used for our own data.  

\begin{figure*}
\centering
\includegraphics[width=12cm]{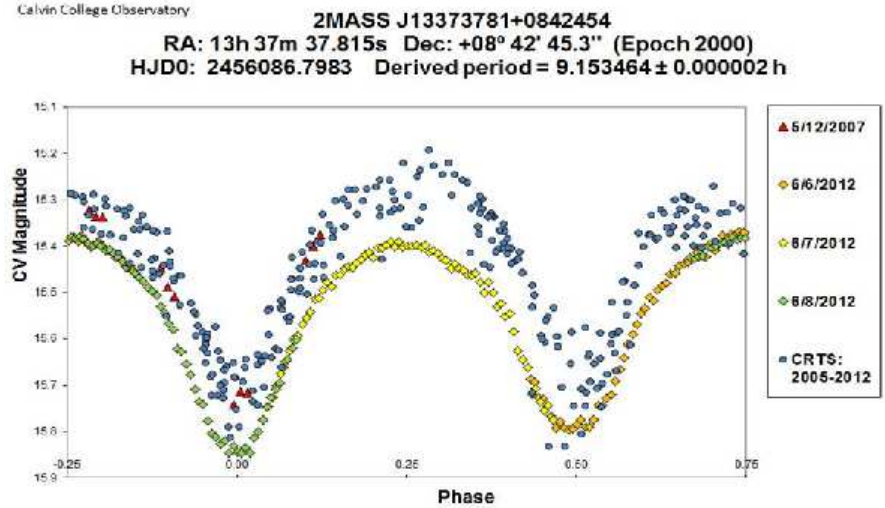}
%\plotone{Steenwyk_Fig3.eps}
\vskip0pt
\caption{Newly discovered contact binary with changing magnitude--first W UMa}
\label{o3}
\end{figure*}

All the data shown in the phase plot are plotted vs. time in Figure 4.  Here the data from each epoch show similar oscillatory amplitudes but their average brightness shows a clear downward trend.  It is reassuring that our 2007 data (red squares) are consistent with the somewhat noisier CRTS data from the same epoch. The earliest epochs suggest the brightness had peaked in 2006 followed by a decrease of about 0.15 magnitudes.  The hint that the brightness may again be increasing seen in the figure is supported by our most recent observations (not shown).    

\begin{figure*}
\centering
\includegraphics[width=12cm]{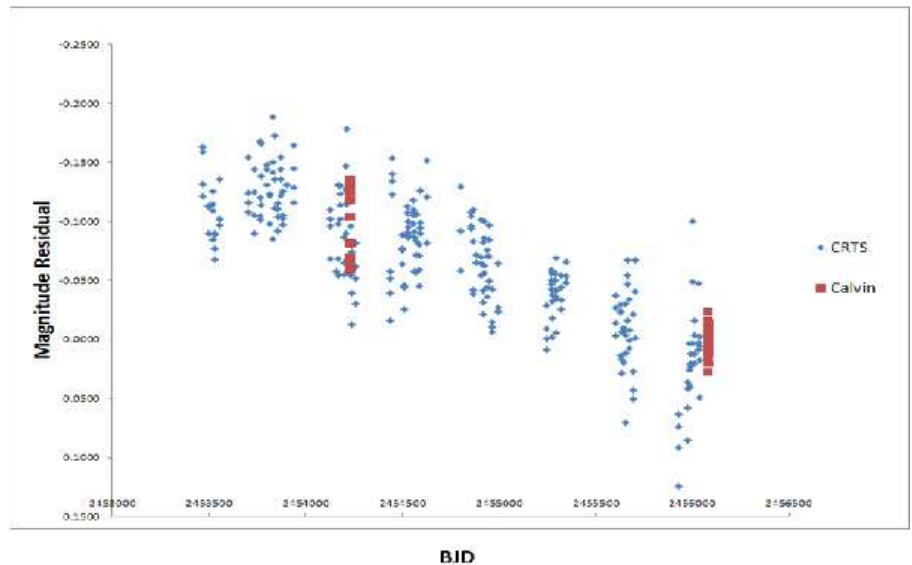}
%\plotone{Steenwyk_Fig4.eps}
\vskip0pt
\caption{Time plot of all observations for J13373781+0842454, first W UMa.}
\label{o4}
\end{figure*}

Careful examination of the 2012 light curve in Figure 3 shows a very slight asymmetry  between phase 0.25 and 0.35 suggesting the presence of a star spot and therefore some degree of magnetic activity.  The slightly lower maximum at phase 0.25 also may suggest a star spot.  Asymmetries attributed to star spots on this scale and even larger are not uncommon in W UMa light curves but do not normally earn designation as an RS system.   Evidence of significant dynamic magnetic activity is required.  Though lacking spectroscopic evidence of chromospheric activity, under the guidelines of the AAVSO's classification system, we classified this star as EW/RS in the VSX due to the slow brightness shift.  It is hard to imagine a coordinated change in star spot coverage distributed over both stars that could account for  the observed uniform shift in brightness.  If star spots are to account for the shift in average magnitude, one would also expect to see localized light curve distortions of similar magnitude.  But nowhere do the light curve data show changing, localized distortions of commensurate magnitude.  

The second W UMa system reported as EW/RS (VSX J062955.8+224813) shows a similar drift in a JD plot of data taken in the Rc passband (Figure 5). This system shows an initial downward shift of 0.08 mag from early 2010 to late 2010 and then a subsequent rise to nearly the initial magnitude in early 2012 so that brightness seems to have passed through a  minimum during our observations. Presumably, we will find a brightness maximum with continued observation. Thus both systems suggest the existence of a cycle of brightness variation on a time scale consistent with stellar magnetic cycles. However, again we saw no concomitant changes in light curve shape to indicate dynamic star spots.  

\begin{figure*}
\centering
\includegraphics[width=12cm]{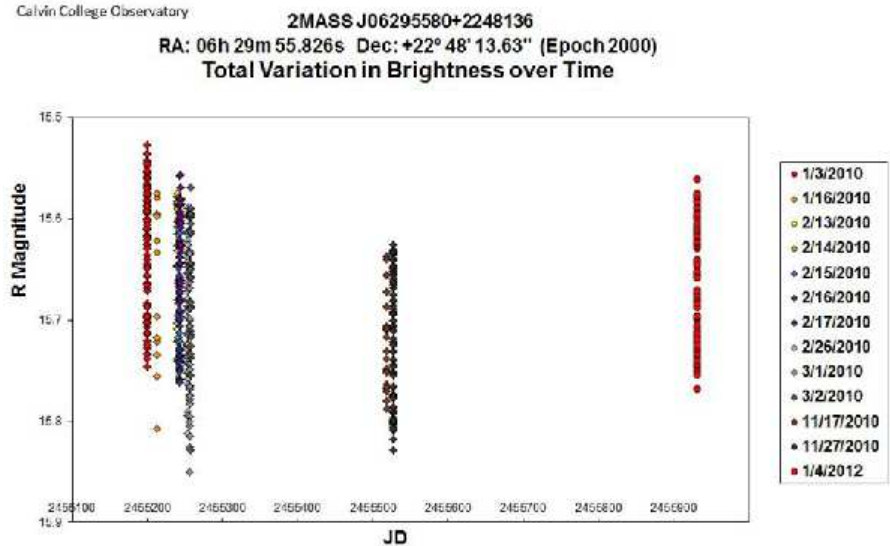}
%\plotone{Steenwyk_Fig5.eps}
\vskip0pt
\caption{Time plot of all observations for second W UMa}
\label{o5}
\end{figure*}

Models have been developed linking magnetic cycles to changes in average brightness on the order of 0.1 magnitude in other chromospherically active stars (Hall 1991).  Some posit a star's rotational inertia changes as magnetic coupling of differentially rotating layers  varies during magnetic cycles that are similar to the 11-year solar cycle.  This may manifest as changing oblateness associated with a redistribution of angular momentum as proposed by Applegate (1992) and reviewed by Ibanoglu (1998).  Whether this provides an adequate explanation of the variations seen in these cases remains to be determined.  The rate of change, especially in the second case, is faster than typical of most of the cases of brightening attributed to this mechanism.  

\section{A Binary in Transition?}
\subsection{The data}
Finally we turn to a most atypical close binary we have found, here designated V0738 (short for VSX J073837.3+295031).  Figure 6 shows differential photometric data from unfiltered observations from three nights in November, 2008.  Variability was discovered in these data in August, 2011, and was submitted as type EB denoting an eclipsing binary showing large ellipsoidal tidal distortion but without direct thermal contact as evidenced by significantly different depths of the minima.  If it were truly in thermal contact via a shared outer envelope, the temperatures of the two bodies would become nearly equal with  minima having nearly equal depth.   Since the submitted light curve was complete and seemingly unambiguous, a follow-up observation was not made until December, 2011 (Figure 7).  We were surprised to find the average brightness had increased by roughly 0.1 magnitude and the minima were more nearly equal. Though it is now clear that the minimum at zero phase was no longer the primary minimum, at the time the data were acquired that was not clear because the period and phasing were not yet established unambiguously.  We also found the peak-to-peak amplitude was greatly reduced, from 0.28 mag at discovery, to 0.14.  An intensive series of follow-up observations found it continued to change.  Though observations from February, 2012, were little changed from three months earlier, observations in March, Figure 8, revealed an even lower amplitude of only 0.07 with minima of nearly equal depth, suggesting stars of equal temperature.  At this point the system had the appearance of a contact binary viewed at a rather low inclination.  Since the inclination could not have changed, we realized this was a highly atypical binary, throwing its original EB classification into doubt.  Had this been the discovery light curve, it would be classified either as EW (W UMa) or ELL (denoting ellipsoidal distortion) in the VSX.

\begin{figure*}
\centering
\includegraphics[width=12cm]{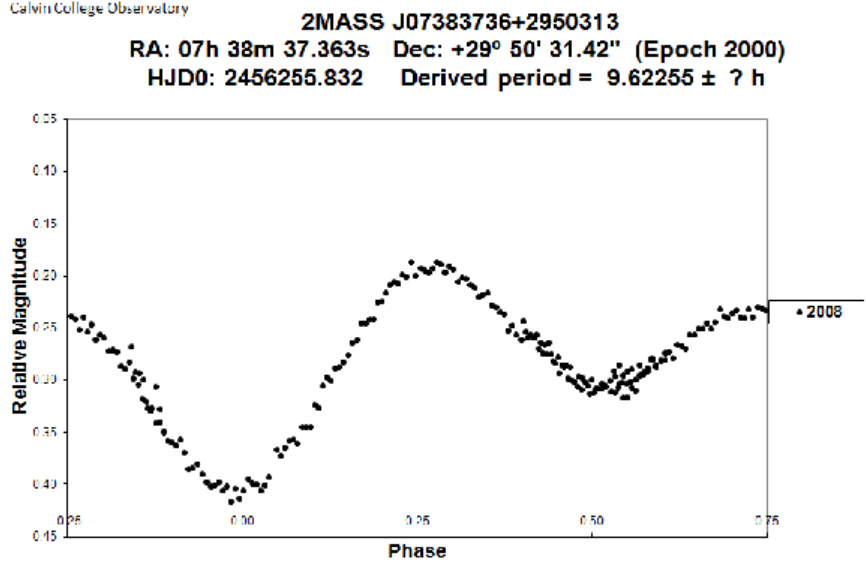}
%\plotone{Steenwyk_Fig6.eps}
\vskip0pt
\caption{2008 discovery data for V0738 showing a detached binary.}
\label{o6}
\end{figure*}

\begin{figure*}
\centering
\plotone{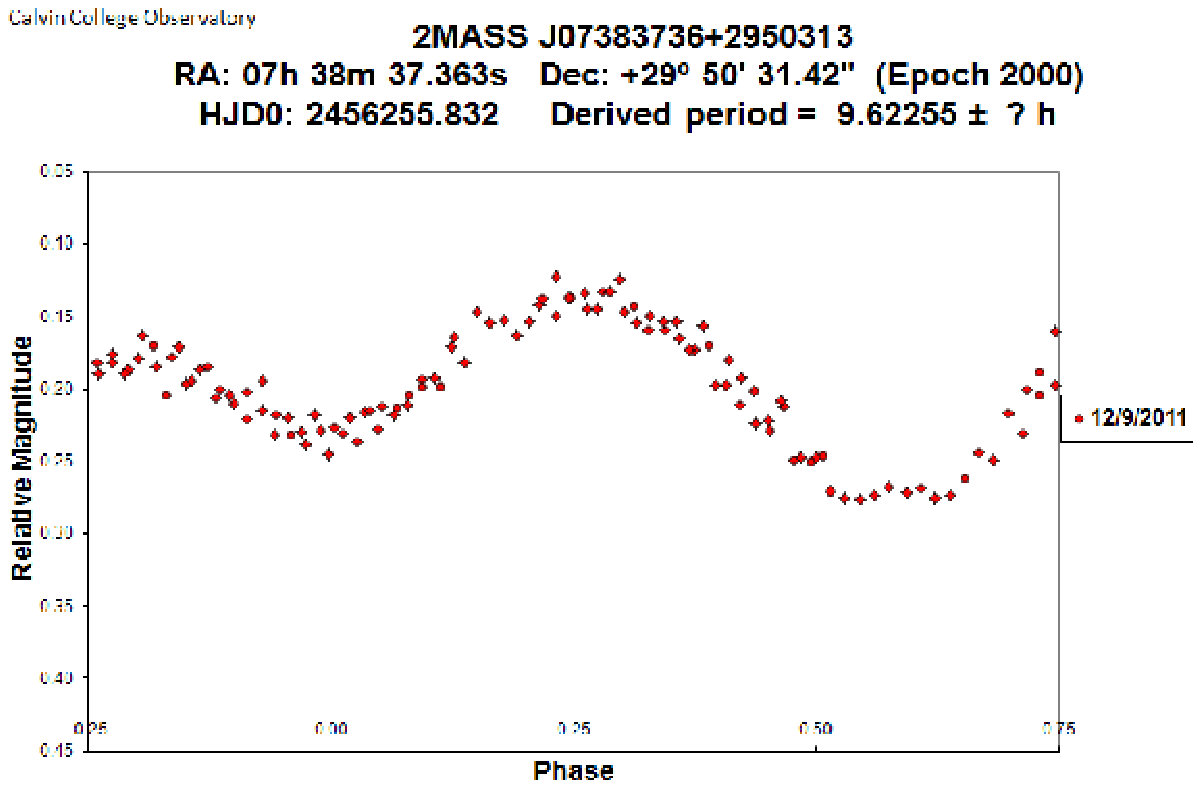}
\vskip0pt
\caption{V0738 in December, 2011, a radical difference: increased luminosity, minima more equal, primary at phase 0.5--a transition has occurred.}
\label{o7}
\end{figure*}

\begin{figure*}
\centering
\includegraphics[width=12cm]{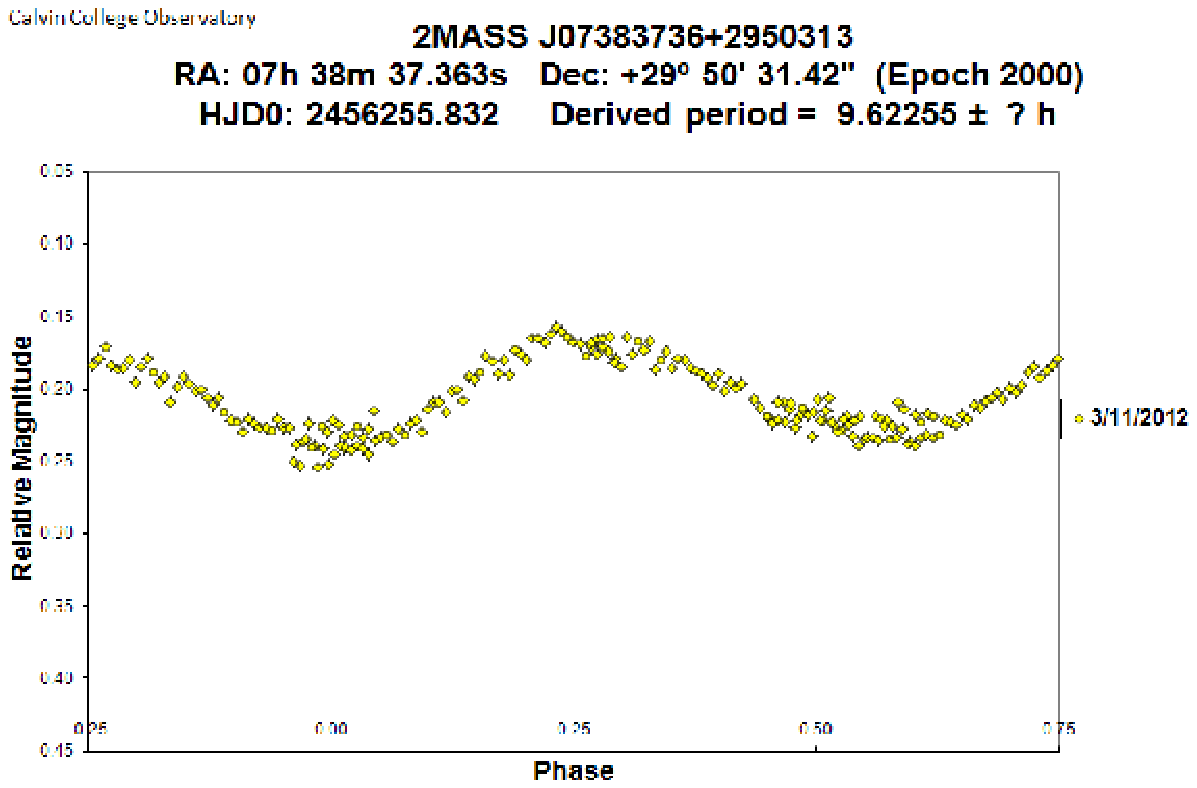}
%\plotone{Steenwyk_Fig8.eps}
\vskip0pt
\caption{V0738 in March, 2012: equal minima suggest equal temperatures, i.e. thermal contact.}
\label{o8}
\end{figure*}

The apparent reversal of the primary and secondary minima bears special elaboration because, as apparent in data taken a year later in December, 2012, (Figure 9) the phase of primary and secondary minima switched again.  Initially, based on only three successive nights of data, the period was reported to only four figures.  Given the intervening three-year interval from the discovery epoch and the shape changes, we were not initially aware that the primary and secondary minima had swapped phases as we were not yet able to establish the phase unambiguously relative to the 2008 epoch.  It was only after many further observations through spring 2012 and a second epoch of many months covering fall 2012 through spring 2013 that we found a single period of 9.62255 hours that reliably links all the data so that we confidently concluded the primary minimum is correctly placed at phase 0.5 in Figure 7 and that it had, in fact, switched back to 0 (as in Figure 9).   Though the exact uncertainty on the final digit of the period cannot be determined due to the shape changes and possible small period changes, these effects are not sufficiently large to cast doubt on the phasing.   The period given is the best average period and our phase error is less than 10\%, eliminating any ambiguity in assigning the correct phase to the minima.

\begin{figure*}
\centering
\includegraphics[width=12cm]{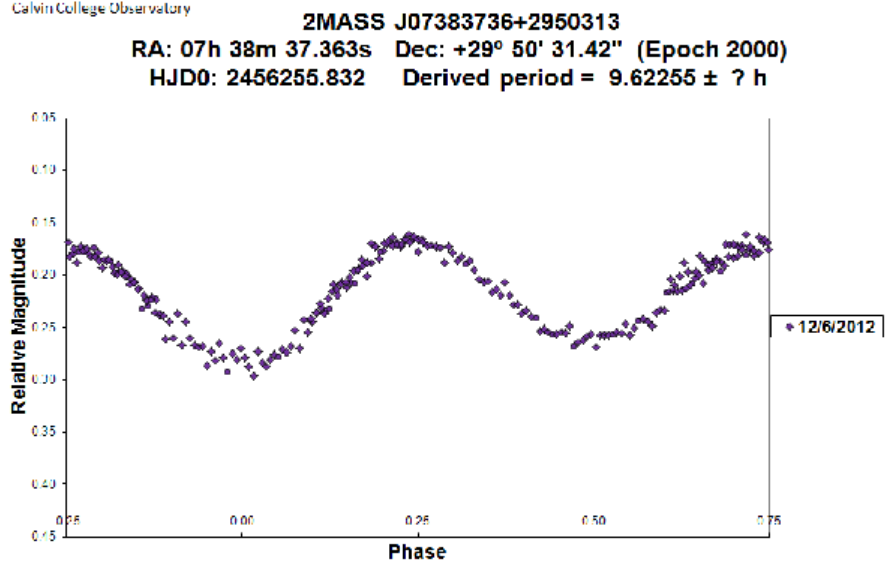}
%\plotone{Steenwyk_Fig9.eps}
\vskip0pt
\caption{V0738 in March, 2012: near equal minima with primary again at phase 0}
\label{o9}
\end{figure*}

Figures 10 and 11 show all data collected from both annual epochs along with the original discovery data from 2008 for reference.  Table 1 summarizes some key features of these changing light curves.  We notice the average magnitude, relative to our reference stars, remained fairly stable, around 0.22, after the initial brightening.  Average magnitude for each month was estimated by eye simply using a ruler to find a line that would produce equal areas above and below the line.  The so-called O'Connell effect is apparent by differing heights of the two maxima. We observed changes in the height of the maxima (not shown in the table) where the difference in maxima ranged from nearly zero to 0.06 magnitude.  However, the more variable maximum at phase 0.25 never was observed to be less than the relatively stable second maximum at phase 0.75.  

\begin{figure*}
\centering
\includegraphics[width=12cm]{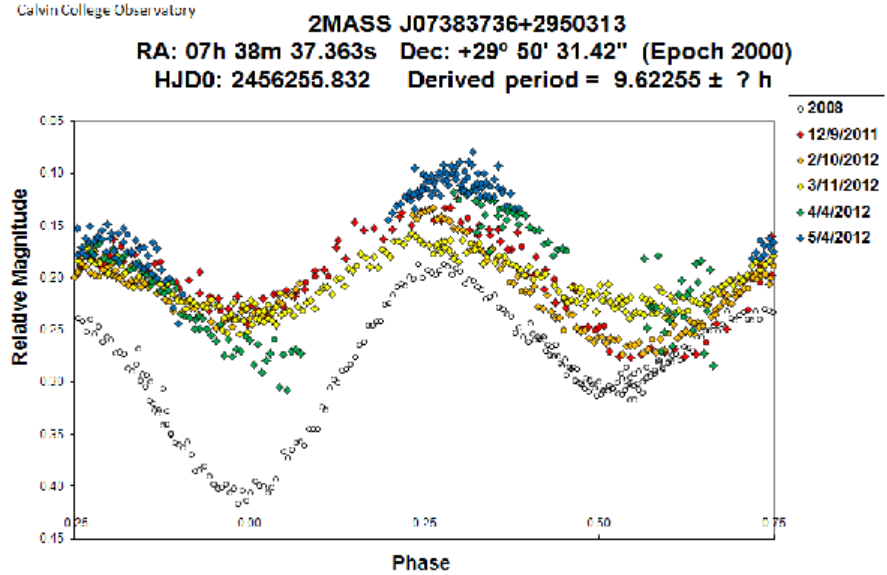}
%\plotone{Steenwyk_Fig10.eps}
\vskip0pt
\caption{All data from the first annual epoch, Dec. 2011 - May, 2012, for V0738. The 2008 discovery data are included for comparison.}
\label{o10}
\end{figure*}

\begin{figure*}
\centering
\includegraphics[width=12cm]{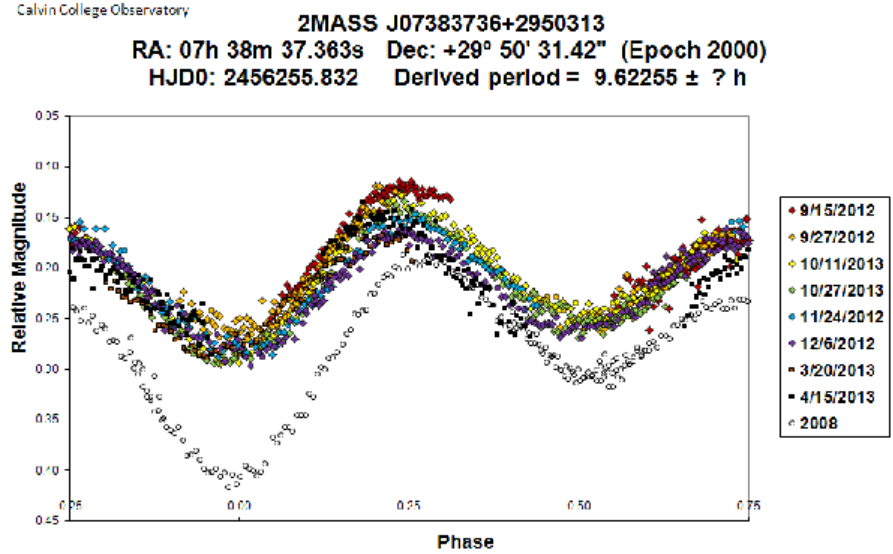}
%\plotone{Steenwyk_Fig11.eps}
\vskip0pt
\caption{All data from the second annual epoch, Sept. 2012 - Apr., 2013, for V0738. The 2008 discovery data are again included for comparison.}
\label{o11}
\end{figure*}

To summarize key points of all of our observations, it may be said that the system showed dynamic changes in the light curve throughout. Though at times it showed relative stability for several months, and other times it changed dramatically from one month to the next. A significant case of dramatic change occurred during the months of February through April, 2012, where we observed a transition  between a state where the primary and secondary minima are reversed to when they are normal, that is, with the primary again at 0 phase as defined by the 2008 data. February data (not shown separately but very similar to Figure 7) clearly show a primary minimum at phase 0.5.  Recall that March, 2012, is the month with very shallow but equal minima (Figure 8). On the other hand, the April, 2012, data (also not shown separately) show minima that had again switched roles back to normal, with the phase zero again being somewhat deeper while the amplitude had again returned to that of February.  Thus it appears a state with low amplitude and equal minima may signal the transition point when the primary and secondary minima swap phases.  The normal phase appears to have continued for nine months through December 2012 (Figure 11).  Then a malfunction at our remote site forced a three-month break in coverage.  When observations were resumed in the spring, it appeared the deeper eclipse was again at phase 0.5. Unfortunately, an opportunity to verify whether the phase switch was accompanied by a state of exceptionally low amplitude was missed.  

%HERE IS TABLE 1

\begin{flushleft}
\begin{deluxetable*}{rcccc}
\tabletypesize{\normalsize}
\tablecaption{Summary of selected features of the evolving light curve of V0738}
\tablewidth{0pt}
\tablehead{ \\ \colhead{Epoch} & \colhead{Est. ave. mag.$^{\rm a}$} &     \colhead{Total amplitude} & \colhead{primary min. ph.} & \colhead{Difference in minima} \\
}

\startdata
 %I--194 &  $-$0.09  & 4210 &  1.9 &  $-$0.1  &  1.04 &  4170\rlap{\tablenotemark{a}} \\
Nov., 2008 &  $0.28\pm 0.01$  & $0.22\pm 0.01$&  0&  0.10 \\
Dec., 2011 &  $0.20\pm 0.01$ & $0.14\pm 0.01$&  0.5&  0.03   \\
Feb., 2012 &  $0.21\pm 0.01$ & $0.13\pm 0.01$ &  0.5 &  0.03   \\
Mar., 2012 &  $0.22 \pm 0.01$  & $0.07 \pm 0.01$&   0.5 &  0.01  \\
Apr., 2012 &  $0.20 \pm 0.01$ & $0.16 \pm 0.01$ & 0 &  0.05  \\
May, 2012 & $0.20 \pm 0.03$  & ...&  ... & ...  \\
Sep., 2012 &  $0.22 \pm 0.01$ & $0.14 \pm 0.01$ & 0 &0.01  \\
Oct., 2012 &   $0.22 \pm 0.01$ & $0.15 \pm 0.01$&  0 &  0.04  \\
Nov., 2012 & $0.22 \pm 0.01$ & $0.13 \pm 0.01$ & 0 &  0.02\rlap{$^{\rm b}$}  \\
Dec., 2012 & $0.22 \pm 0.01$ & $0.11 \pm 0.01$& 0 & 0.02  \\
Mar., 2013 & $0.24 \pm 0.02$ & $>0.10$ & ... & ... \\
Apr., 2013 &  $0.23 \pm 0.02$  &$>0.12\pm 0.02$ &  0.5\rlap{$^{\rm b}$}  & ...  \\
\enddata
\tablenotetext{a}{magnitudes relative to chosen calibration stars}
\tablenotetext{b}{estimated from incomplete light curve when reasonable extrapolation was possible}
\end{deluxetable*}
\end{flushleft}

\subsection{Hints of an explanation}
We now speculate on possible causes of this highly atypical binary.  The time scale of the changes is consistent with the time scale of magnetic activity as seen in RS CVn systems, but the magnitude of changes, reversal of the phase of primary and secondary minima and the apparent transition from non-contact to contact seem, in total, well out of normal behavior for RS CVn types.  We believe this star may be right at the edge of thermal contact, with the stars nearly filling their Roche lobes.  The 2008 observations happened to find the star out of contact, thus showing very different depths of the minima corresponding to different stellar temperatures.   When next observed in December, 2011, it had transitioned to contact.   This would account for the more equal minima.  It would also account for the rise in average brightness, since thermal contact allows the hotter star to acquire a larger radiating surface by adding the secondary star's surface to its own.  A change in oblateness from a magnetically induced change in the angular momentum distribution of one or both stars due to the previously mentioned mechanism of Applegate would provide a mechanism to cause a transition from non-contact to contact in this instance.  

\section{Summary and Conclusions}
%% WE ALL HAD A GOOD TIME, INDEED!!!  THANKS--SDS
Out of 44 W UMas and 18 detached binaries we have submitted to the VSX, only four discussed were actually designated as part of the smaller subset of binaries considered to display RS CVn behavior.  Two displayed the light curve variability expected of evolving star spots on these magnetically active systems.  Three others are not explicable in terms of spots alone, though their behavior may well be associated with magnetic cycles in one or both stars of each pair.  Two display long term shifts in luminosity long associated with RS systems. However, we find it curious that we saw no significant changes in light curve shape one would expect from the dynamic behavior typical of star spots.  We remain open to explanations in terms of magnetic cycles, which seem the most plausible mechanism to act on these time scales.  Finally, we showed an especially intriguing system that may have transitioned from non-contact to contact by a mechanism connected to magnetic cycles.  These cases have certainly made clear that only long term and detailed photometric observations can follow the sometimes rapid changes such systems display.  We intend to continue to follow the systems discussed.  Spectroscopic data would help in further understanding these intriguing binaries.

\section{Acknowledgments}
DMV and SDS acknowledge support from the Michigan Space Grant Consortium.  SDS has also been supported by a Calvin Research Fellowship.  DMV has also received support from the Calvin College Integrated Science Research Institute.


\begin{thebibliography}

\bibitem[Applegate(1992)]{apple} Applegate, J.H. 1992,  ApJ,  385,  621 

\bibitem[Drake et al.(2009)]{drake} Drake, A.J. et al. 2009, ApJ, 696, 870

\bibitem[Hall(1976)]{hall76} Hall, D.S. 1976, “The RS CVn Binaries and Binaries with Similar Properties”, in Multiple Periodic Variable Stars, (Ed.) Fitch, W.S., Proceedings of IAU Colloquium 29, held in Budapest, Hungary, vol 60 of Astrophysics and Space Science Library, 287

\bibitem[Hall(1991)]{hall91} Hall, D. S.  1991, ApJ, 380, L85

\bibitem[Ibanoglu(1999)]{iban} Ibanoglu, C. 1999, Tr.J.Phys, 23, 321

\end{thebibliography}
\end{document}